\documentclass[12pt,preprint]{aastex}

\newcommand \tw {\rm T/|W|}

\begin{document}

\title{De-leptonization and Non--Axisymmetric Instabilities in 
Core Collapse Supernovae}

\author{J. Craig Wheeler, Shizuka Akiyama}
\affil{Department of Astronomy, University of Texas, Austin, TX 78712}
\email{wheel@astro.as.utexas.edu,shizuka@astro.as.utexas.edu}

\begin{abstract}

The timescale of de-leptonization by neutrino loss and associated
contraction of a proto-neutron star is short compared to the time
to progagate a shock through the helium core of a massive star, and 
so the de-leptonization phase does not occur in the vacuum of space, 
but within the supernova ambiance whether or not there has been
a successful explosion. Dynamical non--axisymmetric instabilities 
(NAXI) are predicted for sufficiently strongly differentially 
rotating proto--neutron stars.  Some modes are unstable for small values 
of the ratio of rotational kinetic energy to binding energy, 
${\tw} \gtrsim 0.01$. 
The NAXI are likely to drive magnetoacoustic waves into the 
surrounding time--dependent density structure. These waves 
represent a mechanism of the dissipation of the free energy of 
differential rotation of the proto--neutron star, and the outward 
deposition of this energy may play a role in the supernova explosion 
process. We estimate the power produced by this process and 
the associated timescale and discuss the possible systematics of 
the de-leptonization phase in this context. A likely possibility is 
that the proto-neutron star will spin down through these effects 
before de--leptonization and produce substantial but not excessive 
energy input. 

\end{abstract}


\section{Introduction}

Core collapse of massive stars affects a broad range of astrophysics
including the production of neutron stars and black holes, the creation
and dissemination of heavy elements, and the potential production  of 
gravitational radiation, but the physics of core collapse supernovae 
is not yet fully understood. Neutrino processes play a major role and 
asymmetries are ubiquitous \citep{wan96, wan01, leo06}.  

After its birth as a hot, rapidly spinning proto-neutrons star (PNS) 
at core bounce, a neutron star that avoids being converted to a black 
hole will de--leptonize, shrink in radius, and spin up. The evolution 
in this phase, particularly in the presence of magnetic fields, has 
not been thoroughly explored, but is critical to establish the rotation 
and magnetic field distribution of the neutron star as it becomes 
accessible to observation in the center of the explosion.  
One of the processes that could enhance the depletion of the angular
momentum of the neutron star is the development of non--axisymmetric
instabilities (NAXI) with the attendant generation of magnetoacoustic waves.
When differential rotation is strong, the value of the ratio of 
rotational kinetic energy to binding energy, ${\tw}$, for onset 
of NAXI can be rather low (see \S 2).  If the threshold for NAXI 
is low, these instabilities may be relevant to a significant 
range of rotating core--collapse situations.  Much of the literature 
on NAXI in neutron stars ignores the hot PNS conditions just after core 
collapse, magnetic fields, and the time--dependent density structure 
that surrounds the new--born neutron star until this matter is swept away 
in the supernova explosion.  These complications may have a major impact 
on the rotational and magnetic state of the surviving, observable neutron 
star. We explore some of the possible effects here.

Section 2 gives a brief summary of the literature of non--axisymmetric
instabilities of differentially rotating neutron stars and an estimate of
the rate of dissipation of rotation by NAXI. We consider the 
de--leptonization phase in \S 3.  Conclusions are given in \S 4.

\section{Non--Axisymmetric Modes in Neutron Stars}

For sufficiently large values of ${\tw} \gtrsim 0.27$, rotating 
incompressible spheroids will be dynamically unstable to formation of a
rotating bar configuration \citep{cha69}.  For ${\tw} \gtrsim 0.14$ dissipation 
leads to a secular instability and again to a bar--like mode \citep{cha69, 
sha04, ou04}. For compressible, differentially rotating configurations, 
the limit for dynamical instability can be decreased \citep{toh90, 
ram98, cen01,ima04,shi05}.  \citet{shi03} find that non-axisymmetric 
dynamical instabilities can occur for $\tw$ as small as 0.01. 
This result was obtained in the absence of detailed microphysics that 
might result in damping of the instability, but nevertheless indicates
the possible range of conditions for which such instabilities might occur. 
\citet{wat05} propose that differentially rotating stars have a
continuous spectrum of modes with dynamical behavior that is distinct from
discrete normal modes in uniformly rotating stars.  They explain this 
behavior with the idea of resonances in a ``co--rotation band." 
\citet{ott05} found spiral wave instabilities of $m = 1, 2$ modes 
with $\tw$ = 0.08.  \citet{ot06} explored the growth of these modes and 
their dependence on the presence of a ``cavity" \citep{lov99,li00,li01} 
and co-rotation point. 

Little of the work on non--axisymmetric 
instabilities summarized here includes magnetic fields.  \citet{aki03} 
\citep[see also][]{tho05,wil05,bla06} showed that the MRI is likely to build 
up a strong, primarily toroidal, magnetic field on a time scale of tens 
of milliseconds.  
This conclusion presumes that the shear is not 
damped by other viscosities that would prevent the MRI altogether. 
This assumption must be checked in the future by detailed numerical studies. 
In the extreme case of a collapse that produces 
$\tw \gtrsim 0.20$ at bounce it is likely that a dynamically unstable bar 
will be dissipated by shock and sound waves after several rotations 
and before a strong magnetic field can be generated by the MRI.  
If the initial value of $\tw$ in the PNS is 
$0.14 \lesssim \tw \lesssim 0.20,$  
the magnetic field is expected to grow more rapidly than the 
attendant secular instability. In the context of a PNS,
the dissipation that drives the secular formation of a bar is likely
to be internal viscous dissipation rather than gravitational radiation
reaction forces. An estimate of the dissipation time due to magnetic
shear viscosity by \citet{tho05} gives $\tau_{diss} \sim 
(\alpha \Omega)^{-1}(r/h_P)^2$ where $h_P$ is the pressure scale 
height and $\alpha$ is a dimensionless parameter of order unity.
From their numerical models, \citet{tho05} find $\tau_{diss} \sim 0.1 - 1$
s, much shorter than would be given by either neutrino viscosity
or gravitational radiation reaction \citep{ou04}. 
This estimate, while plausible, does depend on this particular 
$\alpha$ parameter being of order unity. 
The secular bars thus formed would
be Jacobi-like with minimal internal flow, but a rotating pattern
that would interact with the surrounding medium in which the bar rotates.
The general instability is not expected to depend on the manner
in which viscous heat is dissipated, but the specific configuration
may do so \citep{sha04}. 
\citet{ram98} note that a bar--like structure will generate sound and
shock waves like a ``twirling--stick" and stress that collapse is not an
equilibrium situation in terms of the instability and growth of
non--axisymmetric modes.
In what follows we assume that secular bars form by internal viscosity
and dissipate by NAXI more slowly than the magnetic field saturates, 
but much faster than the de-leptonization timescale.   

A saturation time for the magnetic field of tens of milliseconds 
is substantially longer than a dynamical time, but short compared 
to the saturation time of many of the relevant non--axisymmetric 
modes $\sim 50 - 100 $ ms \citep{shi03, ott05, ot06}.  In practice 
the instability to the MRI and those to NAXI may interact in a 
complex way \citep{rez00,rez01a,rez01b}.  That is a very interesting 
issue, but beyond the scope of this paper.  Here we will assume that 
a strong magnetic field does not inhibit the growth of at least some 
of the NAXI, but that the field is established prior to the saturation 
of all the non-axisymmetric modes except the dynamical bar modes.

Another critical time scale in this problem is that for the success of 
a supernova explosion. Given the failure of prompt explosions in 
state of the art model models, the characteristic growth times of 
non--axisymmetric 
modes are long compared to the time for the bounce shock to stall,
10s of ms, but short compared to the time for a successful shock to 
propagate out of the infalling iron core (seconds) and the helium
core ($\sim 10$ seconds) and the time to de--leptonize and contract 
($\sim 1 - 10$ s). If an explosion due to other effects 
(e.g.,  neutrino heating) precedes 
the non--axisymmetric effects, it cannot be very far along when 
the PNS becomes non--axisymmetric. In trial numerical calculations
for which an explosion is induced (to be reported elsewhere), we 
do not find that the density profile is substantially altered
on time scales of several hundred milliseconds, whether or not
an explosion has ensued at that time. The formation and dissipation of 
non--axisymmetric modes in a strongly magnetic PNS thus may 
enhance an explosion that is already underway.

Any non--axisymmetric mode will tend to form sound waves, shock waves and
Alfv\'{e}n and other MHD waves by interaction with the surrounding,
time--dependent, medium. The typical non--axisymmetric mode formed
at modest rotation may be the m = 1 spiral mode discussed by 
\citet{ott05} and \citet{ot06}. While the details will surely
differ, we approximate this spiral-like oscillation by the radial
oscillation of a sphere that has a sinusoidal displacement of
amplitude $\delta r$. Following \citet{ll59}, we write for the 
magneto-acoustic luminosity: 
\begin{equation}
\label{lum1}
L_{mhd} = 2\pi \rho v_{fast} \left(\frac{\delta r}{r}\right)^{2} 
\frac{\omega^4 r^6}{\left(v_{fast}^2 + \omega^2 r^2 \right)},
\end{equation}
where we have taken the velocity amplitude of the oscillations to
be $\delta v = \omega \delta r$ and wave number $k = \omega/v_{fast}$
with $\omega$ the angular frequency of the radiated magnetosonic waves. 
The phase velocity of fast magnetosonic waves is $v_{fast}^2 = 
v_a^2 + c_s^2 \sim c_s^2$ for the case of sub--Keplerian rotation, 
relatively weak saturation fields, and hence $v_a < c_s$. In this 
limit the power radiated into fast magnetosonic waves
and that into ordinary sonic waves is virtually the same.  
The flux into Alfv\'{e}n waves, per se, will be substantially less
than into sound and fast magnetosonic waves.  There are (at least) three
characteristic frequencies involved in this problem, the frequency
of the radiated waves, the frequency of the unstable NAXI, and
the rotational frequency. Instability arguments based on ``co-rotation
bands" suggest that the latter two are comparable. Here we will assume
that all three frequencies are comparable and specifically make
the identification that the frequency of the radiated waves is 
comparable to the rotational frequency, hence $\omega \sim \Omega$.  

With moment of inertia $I = 2/5 f_I M R^2$ and binding energy 
$|W| = 3/5 f_W G M^2/R$ where $f_I$ and $f_W$ are numerical factors of 
order unity we can write \begin{equation}
\label{tw}
\frac{T}{|W|} = \frac{1}{3}\frac{f_I}{f_w}\frac{R^3 \Omega^3}{G M}.
\end{equation}
Eqn. \ref{lum1} can then be expressed as:
\begin{eqnarray}
\label{lum2}
L_{mhd} = \frac{18\pi \rho G^2 M^2}{c_s} \left(\frac{f_w}{f_I} 
  \frac{T}{|W|} \frac{\delta r}{r} \right)^2 
   \left[1 + \left(\frac{v_{rot}}{c_s}\right)^2 \right]^{-1} \\
\nonumber
  \sim 2.5\times10^{56} {\rm erg s^{-1}} \rho_{12} c_{s,9}^{-1} M_{33}^2
  \left(\frac{T}{|W|}\right)^2 \left(\frac{\delta r}{r}\right)^2 
  \left[1 + \left(\frac{v_{rot}}{c_s}\right)^2 \right]^{-1},
\end{eqnarray}
where the subscripts represent normalization in cgs units,
$v_{rot} = \Omega r$ and we expect $v_{rot} \lesssim c_s$.

The timescale for dissipation of the rotational kinetic energy by this MHD
flux is then:
\begin{eqnarray}
\label{tau-mhd}
\tau_{mhd} \sim \frac{\frac{1}{2} I \Omega^2}{L_{mhd}} = 
   \frac{c_s}{30 \pi \rho R G} \left(\frac{f_I^2}{f_W}\right)
   \left(\frac{T}{|W|}\right)^{-1} \left(\frac{\delta r}{r}\right)^{-2} 
   \left[1 + \left(\frac{v_{rot}}{c_s}\right)^2 \right]\\
\nonumber
   \sim 1.6\times10^{-4} {\rm s} \frac{c_{s,9}}{\rho_{12} R_6} 
   \left(\frac{f_I^2}{f_W}\right)
   \left(\frac{T}{|W|}\right)^{-1} \left(\frac{\delta r}{r}\right)^{-2}
   \left[1 + \left(\frac{v_{rot}}{c_s}\right)^2 \right] 
\end{eqnarray}
These results are sensitive to specific parameters, especially the radius 
and the ambient density that will vary with position and time, the
value of $\tw$, and the amplitude of the oscillation. 
Our choice of 
normalization was guided by the density and radius at which the shear 
and magnetic field are maximum and the amplitude by numerical 
simulations \citep{ott05}\footnote{see http://www.aei.mpg.de/~cott/rotinst}. 
With this choice of normalization we find that 
for a PNS of radius 50 km the timescale could be rather short even for
rather small radial perturbations. 
Note that $\tw$ is bounded
by 0.01 and 0.14 in the framework we address here. For $\tw \sim 0.01$,
$R_6 \sim 5$ and $\delta r/r \sim 0.1$, Eqn. \ref{lum2} gives
$L_{mhd} \sim 2.5\times 10^{50}$ erg s$^{-1}$ and Eqn. \ref{tau-mhd} gives 
$\tau_{mhd} \sim 300$ ms, somewhat shorter than the de-leptonization time. 
For the same parameters, but with $\tw$ near the upper limit, the 
luminosity would be 200 times larger and quite competitive with
neutrino heating rates. Note also that these expressions are formally 
independent of the magnetic field (which enters through $v_{fast}$), 
but that for realistic cases with strong gradients in magnetic field 
and perhaps concentrations of the field on the rotational axis, the 
exact nature of the dependence 
on the field strength and distribution could be rather more direct.  
Some of this power could tend to go out the equatorial plane, but 
significant power could be channeled up the rotation axis.

In summary, the ordering of time scales that will affect the physics of
supernovae is roughly:
\begin{equation}
\tau_{dyn} < \tau_{rot} < \tau_{dyn-bar} < \tau_{shock} \sim \tau_{MRI} < 
\tau_{NAXI} \sim \tau_{mhd} \sim \tau_{secular} < \tau_{delep} 
< \tau_{explosion} << \tau_{grr}
\end{equation}
where $\tau_{dyn} \sim 1 $ ms is the dynamical time scale, $\tau_{rot}$ is the 
(sub-Keplerian) rotation timescale, $\tau_{dyn-bar}$, a few rotation times, 
is the timescale for dynamical bar formation, $\tau_{shock} \sim 10$ ms
is the time for the shock to form and stall, $\tau_{MRI} \sim 30 $ ms 
is the time for the MRI to grow the magnetic field to saturation, 
$\tau_{NAXI} \sim 50 -100$ ms is the time for non-axisymmetric (m = 1) modes to grow, 
$\tau_{mhd} \sim 0.1 - 1$ s is the time to spin
down due to magnetosonic luminosity, 
$\tau_{secular} \sim 0.1 - 1$ s is the time for a secular 
bar mode to grow, 
$\tau_{delep} \sim 1 - 10 s$ 
is the de-leptonization time of the PNS to contract to form a neutron star, 
$\tau_{explosion} \sim 10$ s is the time for the successful shock 
to propagate out of the infalling iron core and into the surrounding star,
e. g., the helium core, and $\tau_{grr}$ is the time to dissipate 
angular momentum and rotational energy of the core by gravitational 
radiation reaction forces.  
In the current context, $\tau_{explosion}$ 
is meant to be the time beyond which physical processes in the PNS will 
no longer affect the ultimate outcome, perhaps because the density has 
decreased sufficiently that even if magnetoacoustic flux is liberated, 
it cannot propagate outward and hence affect the explosion. In the
absence of a detailed model of this process, we have taken the time
for a successful shock to propagate into the helium core as a
representative measure of this scale.  We note that the epoch of the
onset of convection within the PNS is of order of tens of ms and
that of convection in the post-shock region is of order 100 ms. 
This emphasizes that timescales of order 0.1 to 1 s, not epochs 
of traditional concentration, will involve a variety of interacting 
physical processes that will need to be more deeply understood. 
We also note that the timescale $\tau_{grr}$ is long, so all the physics 
associated with the other time scales represented here must be solved 
to know the conditions that might be relevant to the production of 
gravity waves once (if ever) those become the dominant sink.

\section{The De--leptonization Phase}

The contraction phase \citep{bur86,kei95,pon99,vil04} will lead to spin up 
and perhaps to crossing the threshold for NAXI or enhancing the growth rate
of these instabilities, the amplitude of which may depend on $\tw$. These 
instabilities will cause some loss of rotation energy and angular 
momentum as the core contracts, perhaps altering the specific 
non--axisymmetric modes that come into play.  The core will dissipate 
its differential rotation and angular momentum until the loss rates 
become comparable to, or longer than, the contraction time scale.  
The de--leptonization phase will depend on details of PNS evolution 
and we will sketch here the qualitative behavior to be expected. 

We schematically summarize various possible behaviors in Fig. 1, 
which presents the loci of evolution in the 
plane representing $\tw$ versus the radius of the contracting PNS 
core, with time as an implicit parameter. Key values of $\tw$, 0.27 
(incompressible dynamical bar), 0.20 (compressible dynamical bar), 
0.14 (secular bar), 0.08 \citep[spiral mode of][]{ott05}, and 0.01 (minimum
for NAXI) are shown as horizontal dashed lines. We also give
on the right axis, an estimate of the effective, mass-weighted
rotation period for a given $\tw$.  We note that these periods 
are only approximate and are, for instance, somewhat longer for 
a given $\tw$ than those given by \citet{ott06}. 

As indicated by the bold arrow on the upper right boundary
of Fig. 1, any PNS born with $\tw \gtrsim 0.14$ is 
likely to spin down quickly to at least $\tw \sim 0.14$ and probably 
to significantly smaller values prior to the contraction phase. 
The radius will, of course, depend on the degree of rotation. 
If the condition of secular instability to a bar mode is reached
by spin-up during contraction, the bar will form, but quickly (in a few 
hundred ms) dissipate its energy to magnetoacoustic waves. 
Stability will be reached, but continued contraction driven by 
neutrino losses will again cause the core to become unstable and 
spin down.  In reality, this will be a continuous process, so that 
if the dissipation by other NAXI modes does not prevent spin up to 
this threshold, then during contraction the core should evolve along 
the locus for secular bar--mode instability, $\tw \sim 0.14$.  
The region above $\tw \sim 0.14$ and to the left of the right hand 
axis is thus a ``forbidden region," which cannot be occupied during the 
contraction phase.  

Thin solid lines in Fig. 1 portray the locus of contraction for 
constant angular momentum, J,  assuming homologous contraction so that 
$\tw \sim J^2/G M^3 R \propto R^{-1}$. For comparison, we also show 
in Fig. 1 the locus of models from \citet{vil04}
who present a series of stationary rotating models with
the thermodynamics set by the evolution of related non-rotating
models. The particular models illustrated in Fig. 1 correspond to
rigidly-rotating configurations for a PNS of 1.6 M$_{\sun}$ rotating
at breakup. The total angular momentum formally decreases along
this sequence. While this is not a fully self-consistent evolutionary
sequence of a rotating, de--leptonizing PNS, it does represent
a possible path of a contracting PNS that sheds angular momentum.
This locus is less steep than $R^{-1}$, presumably reflecting 
the implicit loss of angular momentum.

Schematically, a
PNS born with $\tw \sim 0.08$ that contracts with $\tw \propto R^{-1}$ 
would contact the secular instability line at $\tw \sim 0.14$ and
then follow the locus with $\tw \sim 0.14$ as shown by the horizontal
arrow as the radius continued to shrink until the contraction finished. 
Presuming the structure at that point was still unstable to NAXI,
the evolution would follow the vertical arrow on the left axis. 

Again assuming $\tw \propto R^{-1}$, we find that a PNS born with 
$\tw \sim 0.03$ could contract to $\tw \sim 0.14$ at the end of
the contraction phase as shown by the middle thin line.  Under
the same assumption, a PNS born at the lower limit for NAXI, 
$\tw \sim 0.01$ would contract to $\tw \sim 0.05$, as shown
by the lower thin line. Such a configuration could still be 
unstable to NAXI even after the contraction phase were complete. 
Any PNS born with $\tw \gtrsim 0.01 R_{ns}/R_{pns} \sim 0.002$ might
trigger NAXI before the contraction is complete. Here the subscript 
``pns" refers to the PNS with radius $\sim$ 50 km and the subscript 
``ns" refers to the neutron star with radius $\sim$ 10 km.

A more likely evolution than conserving angular momentum would be 
for the structure to follow the 
locus defined by the equivalence of the contraction timescale and 
the time scale for loss of differential rotation free energy to 
magnetoacoustic waves excited by NAXI.  Deducing this
locus will require elaborate calculations. Based on our
estimate of the timescale for dissipation of the rotational
free energy given in Eqn. \ref{tau-mhd}, our best guess is that
the dissipation time for NAXI is initially rapid compared to
the contraction time. This would suggest an evolution to
lower $\tw$ at essentially constant radius, as shown by
the lower vertical arrow on the right side of Fig. 1. As
the radius decreases, the timescale for dissipation will
lengthen. The amplitude of the perturbation is also likely
to decrease as $\tw$ declines. As the dissipation time
lengthens to be comparable to the contraction time, the
locus of evolution will move to the left in Fig. 1.  In Fig. 1, we
show one possibility, an evolution parallel to, but somewhat
above the minimum value of $\tw \sim 0.01$, the notion
again being that if contraction increases $\tw$ too much,
the dissipation will increase to keep the timescales 
approximately comparable.  

Once full contraction is achieved, the contracted neutron star is 
still likely to be spinning faster than any infalling or outgoing 
material around it, so there could also be a final spin-down phase 
with the neutron star at constant radius until the inner structure 
were completely stable to NAXI and other interactions. A rotating
neutrino wind might contribute at this epoch \citep{tho04}.

A key question is how much energy can be liberated as
magnetoacoustic flux during this contraction phase. 
Most of the difference in binding energy between the PNS and 
the final, contracted, cool, neutron star will be emitted in neutrinos 
(although with steadily smaller luminosity), but for a rotating, 
magnetized PNS, some of this energy will be dissipated in 
magnetoacoustic power. In one extreme case corresponding to 
continued rapid (on the contraction timescale) dissipation by NAXI, 
the initial rotational energy would be dissipated before contraction 
would occur. In that case, virtually all the excess binding energy
would be lost to neutrinos, and the available rotational
energy would just be the initial value of the free energy in 
differential rotation.  

In the opposite extreme case that angular
momentum is conserved during the contraction phase, J = const, 
which is equivalent to no dissipation into magnetoacoustic flux, we have
after contraction, but before final spin-down:
\begin{equation}
\label{jconst}
T_{ns} \sim \frac{W_{ns}}{W_{pns}}\frac{R_{pns}}{R_{ns}}T_{pns} \sim 
\left(\frac{R_{pns}}{R_{ns}}\right)^2 T_{pns}.
\end{equation}
We have again assumed that $\tw$ and $W$ both scale as $R^{-1}$.
In this case, a maximal amount of the excess binding energy 
is converted to rotational energy after contraction.  This condition of 
constant angular momentum is commonly assumed in contraction of isolated 
neutron stars \citep{vil04,ott06}, but is unlikely to apply to real 
rotating magnetic PNS still buried within the collapse ambiance
unless they are rotating very slowly.  

If the PNS evolves by shedding magnetoacoustic energy at roughly
constant $\tw$ then
\begin{equation}
\label{twconst}
T_{ns} \sim \frac{W_{ns}}{W_{pns}} T_{pns} \sim \frac{R_{pns}}{R_{ns}}T_{pns},
\end{equation}
and
\begin{equation}
J_{ns} \sim \sqrt{\frac{T_{ns}}{T_{pns}}}\frac{R_{ns}}{R_{pns}} J_{pns} 
\sim \sqrt{\frac{R_{ns}}{R_{pns}}} J_{pns}.
\end{equation} 
These expressions again assume that the contraction is homologous so that 
the change in radius is an appropriate measure of the change in binding
energy.  
One example of such possible behavior would be if the PNS 
were born or quickly evolved to a condition of marginal stability 
to a secular bar mode and then contraction occurs along the 
locus of secular bar instability with $\tw \sim 0.14$. Another
example would be contraction along the locus near the lower
threshold for exciting NAXI, as illustrated in Fig. 1. 

The amount of energy that could be dumped from the rotation to 
magnetoacoustic energy during the contraction is thus a rather 
sensitive function of how the dissipation affects the angular 
momentum. In the extreme case of conserved angular momentum, if the 
contraction were quasi--homologous so that $\tw$ did scale closely 
as $R^{-1}$, then Eqn. \ref{jconst} suggests that a fraction 
$(R_{pns}/R_{ns})(T_{pns}/|W_{pns}|)$ of the final binding energy, 
something of order 25 times the initial rotational energy, could be 
invested in rotational energy and then liberated in spin down 
magnetoacoustic power.  This rotational energy must be limited since, 
as we argue here, no contracting configuration can penetrate into the 
forbidden zone at $\tw \gtrsim 0.14$. As an example, if we assume 
that the PNS is born with $\tw \sim 0.08$ and contracts along a locus
of J = const (uppermost thin line in Fig. 1) with negligible production
of magnetoacoustic flux until it reaches $\tw \sim 0.14$ and then 
follows that locus during the remainder of the contraction, 
the configuration will end up at the end of contraction with
$\tw \sim 0.14$ and hence $T_{ns} \sim 0.70 |W_{pns}| \sim 9 T_{pns}$,
assuming the radius has contracted by a factor of 5 and the
binding energy has increased proportionately. This energy, still
a considerable amount but much less than 25$T_{pns}$, would then 
be available in a final spin down phase.

Note that this estimate represents a lower limit to the total amount of 
magnetoacoustic power liberated because some must be emitted during 
the contraction along the locus $\tw \sim 0.14$, the very reason that 
locus is followed. The virial theorem gives a constraint on the
total energy lost to neutrinos and magnetoacoustic flux:
\begin{equation}
\Delta E_{\nu} + \Delta E_{NAXI} = - \left[\frac{1-2\eta}{3<\gamma -1>} 
   + \eta-1\right]\Delta W,
\end{equation}
where $\eta$ is the ratio of $\tw$ along the locus, assumed to be constant, 
$<\gamma -1>$ is an appropriate average over the adiabatic index,
and $\Delta W$ is the (negative) change in the binding energy. 
This gives $\Delta E_{\nu} + \Delta E_{NAXI} \sim \Delta W \sim W_{ns}$, 
but the fraction of the energy lost that goes to neutrinos and that to
magnetoacoustic flux would have to be determined by a direct integration
of the appropriate luminosity over the contraction time.  The loss to 
neutrinos is likely to dominate the loss to magnetoacoustic flux. 
  
Contraction from $\tw \sim 0.03$ with constant J (middle thin
line in Fig. 1) would, in this illustration, reach $\tw \sim 0.14$
just as contraction was complete. This would give a rotational
energy at that point to be dissipated of $T_{ns} \sim 0.7 W_{pns}$. 
For initial $\tw \sim 0.01$, the smallest values in the literature 
associated with NAXI, the spin energy available to be liberated 
after contraction if angular momentum were conserved during contraction 
could be $\sim 0.05 |W_{pns}| \sim 0.25 |W_{pns}|$.  

As argued above, the rapid timescale for the dissipation of NAXI
(Eqn. \ref{tau-mhd}) suggests that much of the rotational dissipation
will occur before contraction. In this case, the initial rotational 
free energy will be dissipated before any contraction occurs 
and $\Delta T \sim T_{pns} \lesssim 0.14 W_{pns}$. As an example, if a
PNS is born with $\tw \sim 0.08$, it will lose rotational energy
$T \sim 0.08 W_{pns}$ in prompt spin down. Subsequently, as illustrated
by the lower horizontal arrow in Fig. 1, contraction could drive
the structure along the locus with $\tw \gtrsim 0.01$ with relatively 
little more energy in magnetoacoustic flux.  The actual evolution is 
likely to fall between the extremes of conservation of angular momentum 
during contraction and rapid spin-down prior to contraction.

\section{Conclusions} 

The implication is that if a supernova does not succeed on a time scale of
$\sim 100$ ms, de--leptonization is likely to cause the PNS to contract and
spin up and amplify the driving and magnetic dissipation of
non--axisymmetric modes that could pump energy into the environment of 
the neutron star, potentially energizing the explosion if 
$\tw \gtrsim 0.01$ during the evolution.  Even if an 
explosion has been launched on a time scale of $\sim 100$ ms, the infall 
will not have declined to zero in the early part of the contraction process.  
The ambient density will still give a medium in which to produce and 
dissipate magnetoacoustic waves generated by non--axisymmetric instabilities. 

The contraction phase could thus be important for both the energetics 
of the explosion and the rotational period of the contracted neutron 
star, so it is important 
to understand this phase. We have made some estimates of the energy
that could be deposited by magnetoacoustic waves, but caution that
this will depend on the actual free energy of differential rotation,
the details of the change of structure and hence binding energy
during contraction and the mode of the magnetoacoustic flux. If
the latter were to emerge up the axis as Poynting flux, it might
not contribute to the kinetic energy budget of the expanding ejecta.
The contraction could lead to a rapidly rotating neutron star.
As we show here, however, a plausible result is a rather rapid spin
down prior to substantial contraction followed by contraction at roughly
constant $\tw \sim 0.01$ corresponding to the threshold for NAXI.
That would correspond to a post-contraction rotational period of 
$\sim$ 3 ms prior to any final spin down.

The evolution during the de--leptonization phase will depend on
the physics and time scales of the dissipation of the free energy
of differential rotation. We noted that in our simple model for
the time scale associated with the magnetoacoustic flux generated by 
NAXI, Eqn. \ref{tau-mhd}, the time scale could be rather short 
compared to the contraction time.  Eqn. \ref{tau-mhd} depends on the 
rotation frequency implicitly through
$\tw \propto \Omega^2$ and $\delta r/r$.  This is in contrast to 
the dissipation timescale in a model in which the magnetic field serves 
to provide a magnetic viscosity \citep{tho05,wil05,ott06}, for which 
the dissipation power scales as $\Omega^{3}$ and hence the time scale
as $\Omega^{-1}$.  We note that the time scale
associated with this process is still model dependent with
\citet{tho05} and \citet{ott06} adopting a dissipation power per unit mass of 
$\sim \alpha h_P^2\Omega(d\Omega/d ln r)^2 \sim \alpha h_P^2\Omega^3$
and hence a time scale of $\tau_{diss} \sim (\alpha \Omega)^{-1}(r/h_P)^2$.
\citet{wil05} invoke a threshold magnetic field but effectively adopt 
a dissipation power per unit mass of $1/2 \Omega^3 r^2$ and     
hence a time scale of $\sim \Omega^{-1}$. While both of these time scales
vary as $\Omega^{-1}$, that of \citet{wil05} is shorter by a factor
of $\sim (h_P/r)^2$. 

Another important difference between the
dissipation by NAXI and magnetoacoustic flux and dissipation by
magnetic viscosity is that the magnetic viscosity is, by assumption,
dissipated locally as heat, whereas the magnetoacoustic flux may,
and probably will, propagate and deposit the dissipated energy
non-locally. Deposition near the standing shock, if one remains,
is a likely locale for dissipation, since the magnetoacoustic waves 
are likely to propagate down the density gradient and unlikely to 
penetrate the entropy jump at the shock. Determining the fraction
of the free energy of differential rotation that is dissipated by
magnetic or other forms of viscosity as a local heat source
and the fraction radiated and deposited non-locally in 
magnetoacoustic flux will require detailed calculations.  

The contraction phase of a PNS within the context of realistic 
supernova conditions (successful or not) has not been considered 
in the literature, but in the context of rotating, magnetic, 
non--axisymmetric structures, this phase may be critical to 
understand the energetics of the explosion and the post--supernova 
state of the neutron star.  The contraction phase, however it behaves, 
will be complete before any neutron star is exposed to external observations. 

\acknowledgments
We thank Pawan Kumar, Joel Tohline, Adam Burrows, Tony Mezzacappa, 
John Blondin, Eric Blackman, Jose Pons, and Loic Villain for useful 
conversations.  This work was supported in part by NASA Grant NNG04GL00G 
and NSF Grant AST--0406740.

\clearpage

\begin{figure}
\centering
\includegraphics[totalheight=0.8\textheight]{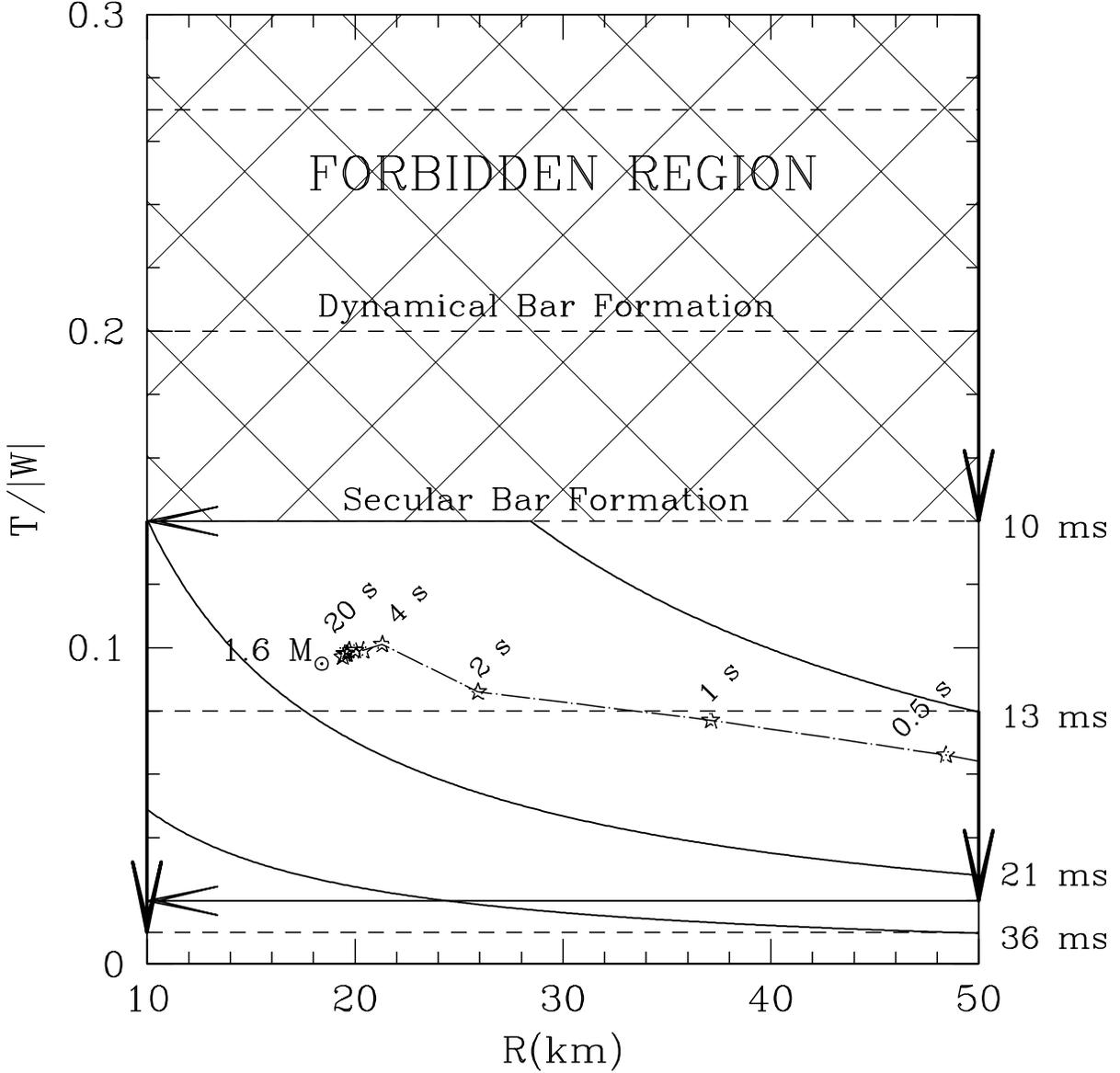}
\caption[contract.eps]{Characteristics of the de-leptonization, 
contraction phase of rotating, magnetic neutron stars are shown
in the plane of $\tw$ versus R. Several key values of $\tw$
representing thresholds for instability (or other representative
values) are given by dashed horizontal lines. The curved lines 
represent homologous contraction conserving angular momentum. Arrows 
represent possible loci of contraction under the influence of 
non-axisymmetric instabilites and radiation of magnetoacoustic
flux (see text for details). The curve denoted by time marks
is taken from \citet{vil04}. The right--hand axis gives an estimate
of the rotational period for a given value of $\tw$ based on a simple
model with constant density within the inner solar mass.  
\label{contract}}
\end{figure}

\clearpage


\end{document}